\documentclass[3p,times]{elsarticle}

\usepackage{ecrc}

\usepackage{epstopdf,here}

\volume{00}

\firstpage{1}

\journalname{Nuclear Physics A}

\runauth{P.B. Gossiaux et al.}

\jid{nupha}

\jnltitlelogo{Nuclear Physics A}

\usepackage{graphicx}
\usepackage{amsmath,amssymb}
\begin{document}

\begin{frontmatter}



\title{Gluon radiation by heavy quarks at intermediate energies and consequences for the
mass hierarchy of energy loss}

%% Single author (and collaboration) - please insert
\author[label1]{P. B. Gossiaux \corref{cor1}}
\author[label1]{J. Aichelin}
\author[label1]{Th. Gousset}
\author[label2]{M. Nahrgang}
\author[label1]{V. Ozvenchuk}
\author[label1]{K. Werner}%\fntext[col1] {A list of members of the XYZ Collaboration and acknowledgements can be found at the end of this issue.}
\address[label1]{SUBATECH, UMR 6457, Universit\'e de Nantes, Ecole des Mines de Nantes, IN2P3/CNRS. 4 rue Alfred Kastler, 44307 Nantes cedex 3, France}
\address[label2]{Department of Physics, Duke University, Durham, North Carolina 27708-0305, USA}

\cortext[cor1]{Corresponding author}
%% For multiple authors, replace the above by:

%\author[label1]{Author1}
%\author[label2]{Author2}

%\address[label1]{Address 1}
%\address[label2]{Address 2}

\begin{abstract}
We extend the Gunion Bertsch calculation of gluon radiation in single scattering to the case of finite mass quarks. 
This case applies to the radiative energy-loss of heavy quarks of intermediate energies propagating in a quark 
gluon plasma. We discuss more specifically the dead cone effect as well as the mass hierarchy of the 
collisional and radiative energy loss and provide some predictions for observables sensitive to the mass hierarchy
of energy loss in ultrarelativistic heavy ion collisions.
\end{abstract}


\end{frontmatter}

% \linenumbers


\section{Introduction}
\label{intro}
The quenching of heavy quarks (HQ) produced in the initial stage of ultrarelativistic heavy ion  
collisions is generally accepted as a good probe of the quark-gluon plasma created in those 
collisions. Among other interesting features, one usually quotes the mass hierarchy
expected in the energy loss and thermalization processes, which offers a different perspective as compared
to the quenching of light particles. This mass hierarchy is often discussed intricately 
with the dead cone effect which reduces the gluon bremsstrahlung in HQ production and could
affect the {\it induced} radiation in QGP as well~\cite{DK}. One goal of this contribution 
is to show that, strictly speaking, the dead cone effect is not the main contributor to the mass 
hierarchy found in radiative energy-loss. For this purpose, we rely on our recent calculation of the radiative 
energy-loss of heavy quarks at intermediate energies~\cite{AGG}, summarized in section \ref{sect_rad_eloss_interm}. 
In such a regime, the typical invariant square 
mass $s$ of collisions with the QGP constituents is not $\gg m_Q^2$ and the coherence effects are not expected 
to dominate the physics, but rather the phase space boundaries.
We have therefore advocated in~\cite{AGG} that extending the calculation of Gunion and Bertsch (GB)~\cite{GB} to finite 
quark-mass could offer a valid alternative perspective to the "high energy approach"~\cite{HE}, especially
for the case of identified open-beauty mesons produced in AA collisions.

Another goal of this contribution is to remind the reader that the mass hierarchy found in radiative energy loss is 
also inherent to collisional energy-loss, as shown in section \ref{massdepEloss} where various models are
compared. This implies that recent experimental evidence of such mass hierarchy in heavy-flavour quenching~\cite{alice} 
has to be considered with care before precise conclusions can be drawn on its origin. In order to contribute to this 
topic, we have implemented, in section  \ref{conseq_URICH}, different energy-loss models in our dynamical MC$@_s$HQ+EPOS2
simulator~\cite{Nahrgang1} and in order to make predictions for observables sensitive to their mass dependence.

\section{Radiative energy loss at intermediate energy}
\label{sect_rad_eloss_interm}
We summarize the main points of \cite{AGG}, in which the reader can find further details and 
explanations. For intermediate energies, coherence effects can be neglected and the dominant contribution to the radiated gluon
stems from a gauge-invariant ensemble of diagrams, including the one where the radiated gluon is attached to the 
gluon exchanged between the light parton (of 4-momentum $q$) and the HQ (of 4-momentum $P$). 
A compact and exact expression is found for the dominant part of the radiation probability:
\begin{equation}
\frac{d\sigma^{Qq\rightarrow Qqg}}{dx d^2k_t d^2l_t}=
\frac{\Theta(\Delta)}{8(2\pi)^5(s-m_Q^2)\sqrt{\Delta}}\,\left|
g\,C_3\,\left(\frac{-4\,g^2\,P\cdot q}{\ell^2}\right)
\left(\frac{(2(1-x)-x')\,\vec{\epsilon}_t\cdot\vec{k}_t}{\vec{k}_t^2+x^2 m_Q^2}
-\frac{2(1-x-x')\,\vec{\epsilon}_t\cdot(\vec{k}_t-\vec{l}_t)}
{(\vec{k}_t-\vec{l}_t)^2 +(x+x')^2 m_Q^2}\right)\right|^2,
\label{eq:dsig_exact}
\end{equation}
where $\vec{k}_t$ and $\vec{l}_t$ are respectively the transverse momentum of the radiated and of the exchanged gluons 
(with respect to the incoming HQ direction), $x$ is the fraction of $P$ carried away by the gluon (after Sudakoff decomposition)
and $\Delta$ represents the phase space. When $k_t^2,l_t^2\ll x(s-m_Q^2)$, eq.~(\ref{eq:dsig_exact}) simplifies and one recovers the 
"high energy limit":
\begin{equation}
\frac{d\sigma^{Qq\rightarrow Qqg}}{dx d^2k_t d^2l_t}=
\frac{d\sigma_{\rm el}}{d^2l_t} P_g(x,\vec{k}_t,\vec{l}_t)
\quad\text{with}\quad 
P_g=\frac{C_A \alpha_s}{\pi^2}
\left(\frac{\vec{k}_t}{k_t^2+x^2 m_Q^2}-\frac{\vec{k}_t-\vec{l}_t}{(\vec{k}_t-\vec{l}_t)^2+x^2m_Q^2}
\right)^2\,,
\label{eq:HE_limit}
\end{equation} 
There exists 2 clear-cut regimes: In the "hard scattering regime" $l_t\gg x m_Q$ 
($\Leftrightarrow x\ll x_M=\frac{\mu}{m_Q}$, where $\mu$ is the IR regulator introduced in 
$\frac{d\sigma_{\rm el}}{d^2l_t}$), one has $P_g\approx \left(\frac{k_t^2}{k_t^2+x^2 m_Q^2}\right)^2$,
which is the usual dead cone effect advocated in~\cite{DK}. In the "soft scattering regime" $l_t\ll x m_Q$,
both terms in $P_g$ interfere and the dead cone in $k_t$ space is replaced by a plateau of height 
$P_g\propto\frac{l_t^2}{(x m_Q)^4}$. This effect has first been shown in \cite{Kaempfer}. 
\begin{figure}[H]
\begin{center}
\includegraphics*[width=8.cm]{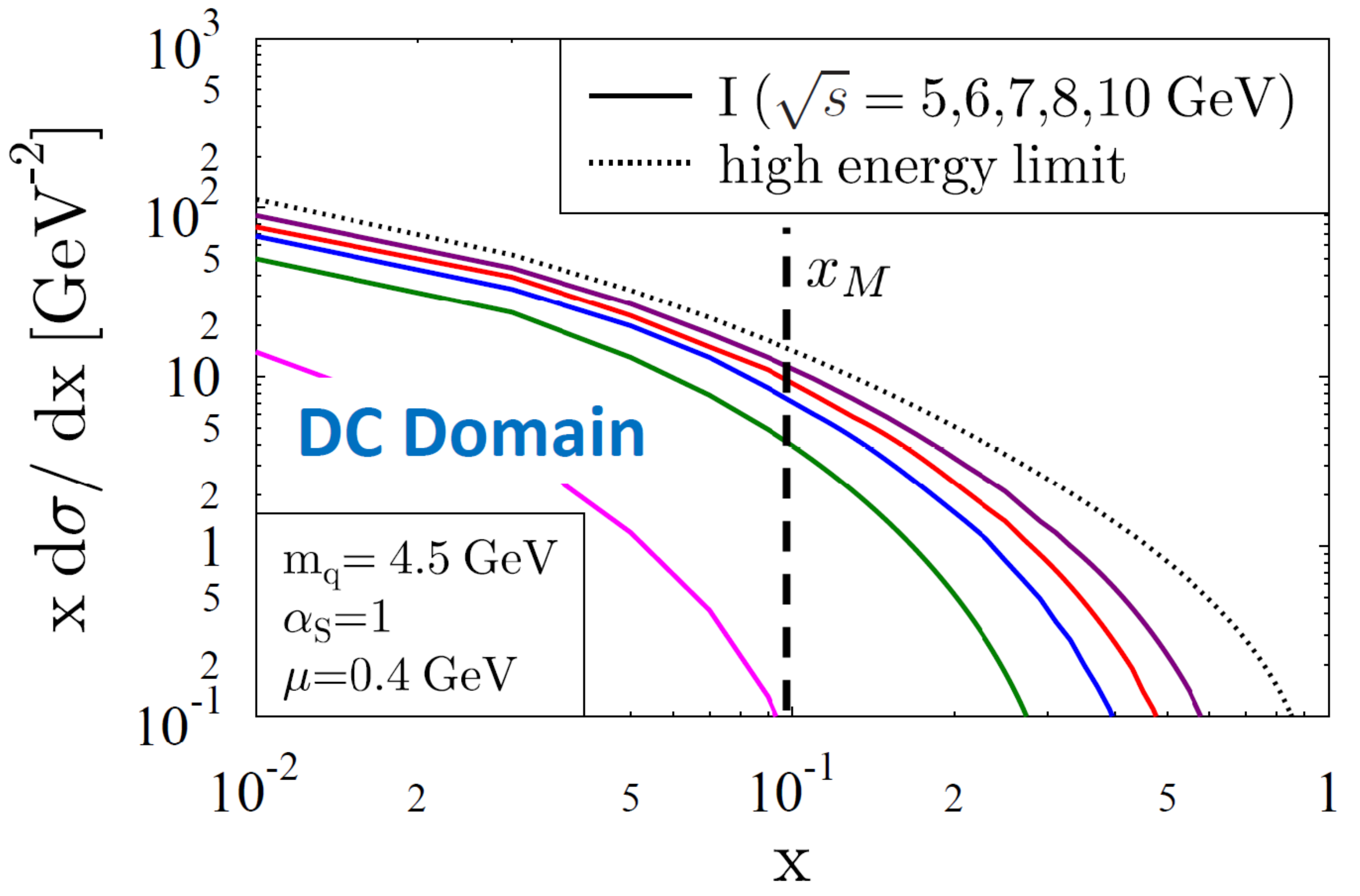}
\includegraphics*[width=8.cm]{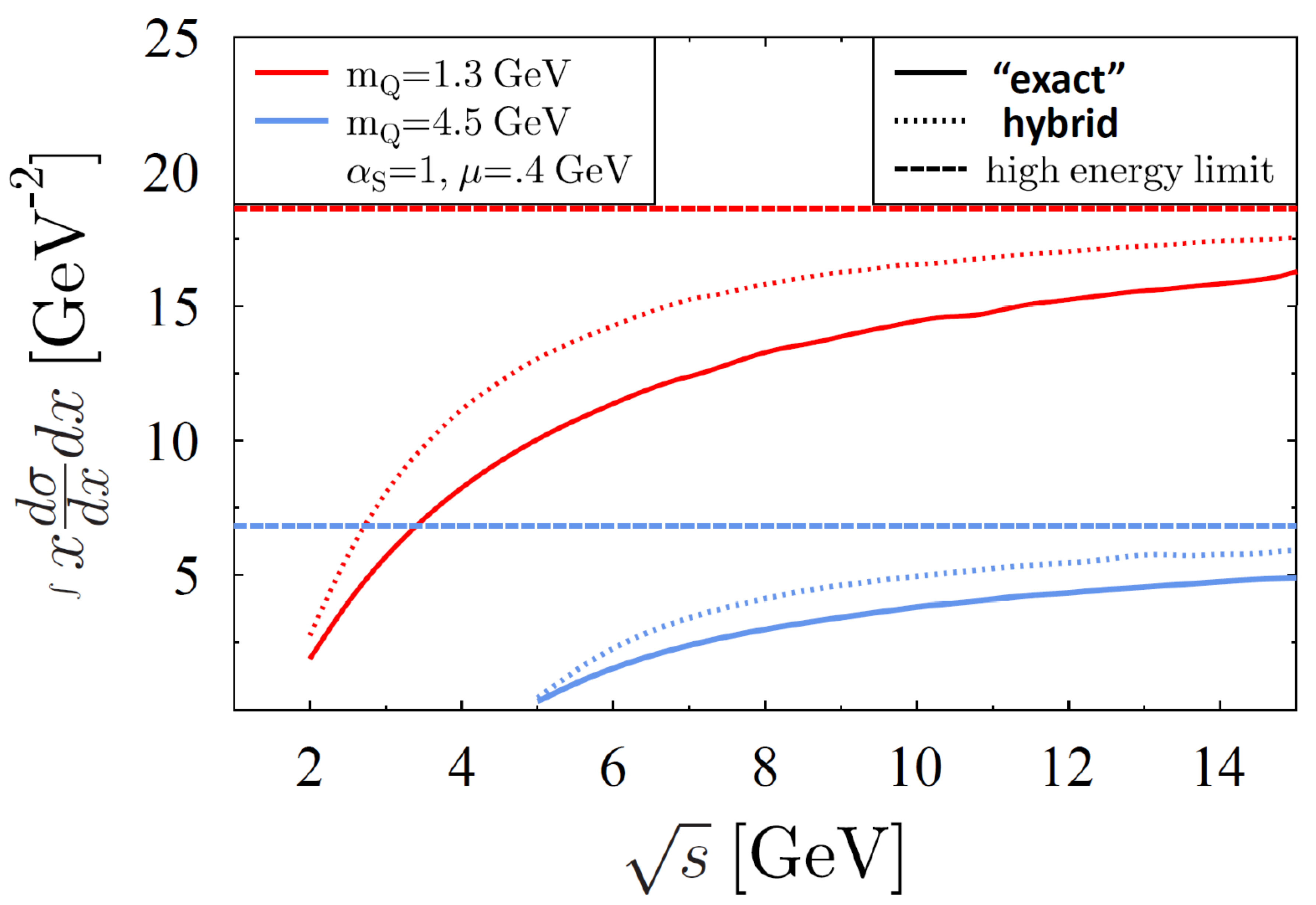}
\caption{Left: the gluon emission cross section $x \frac{d\sigma}{dx}$ in a $b$-$q$ radiative collision. Right: the
moment $\int x\frac{d\sigma}{dx} dx$.} 
\label{fig:rad_elos}
\end{center}
\end{figure}
In fig.~\ref{fig:rad_elos} (left), we display the gluon emission cross section for realistic values of $\sqrt{s}$ 
in a collision  between a $b$-quark of 10 GeV/c and partons of a $T=0.4$ GeV QGP, as well as for the HE limit 
(\ref{eq:HE_limit}) which constitutes an upper value. On the right panel, we show the moment 
$\mathcal{I}=\int x\frac{d\sigma}{dx} dx$ appearing in the average Eloss as $\frac{dE_{\rm rad}}{dz}=\mathcal{I}\rho E$. 
$\mathcal{I}$ is $\sqrt{s}$-independent in the HE limit but strongly reduced at intermediate energies. Moreover, $\mathcal{I}$ scales 
$\propto m_Q^{-1}$ when plotted as a function of $\sqrt{s}-m_Q$. On fig.~\ref{fig:rad_elos} (right), we also show the 
result of the "hybrid" model obtained by combining the $d\sigma$ in eq.~(\ref{eq:HE_limit}) with an
exact phase space boundary. This hybrid model already contains the main features of the exact results but is much more efficient
for MC implementation. It was therefore chosen for our code, after including a gluon thermal mass $m_g=2T$ for the radiated gluon. 
We model coherence effects of the LPM type according to~\cite{gossiaux2012}, which has an impact for $p_T\gtrsim 10~{\rm GeV/c}$
only.

\section{mass dependence of collisional an radiative energy loss models}
\label{massdepEloss}

\begin{figure}[H]
\begin{center}
\includegraphics*[width=7.cm]{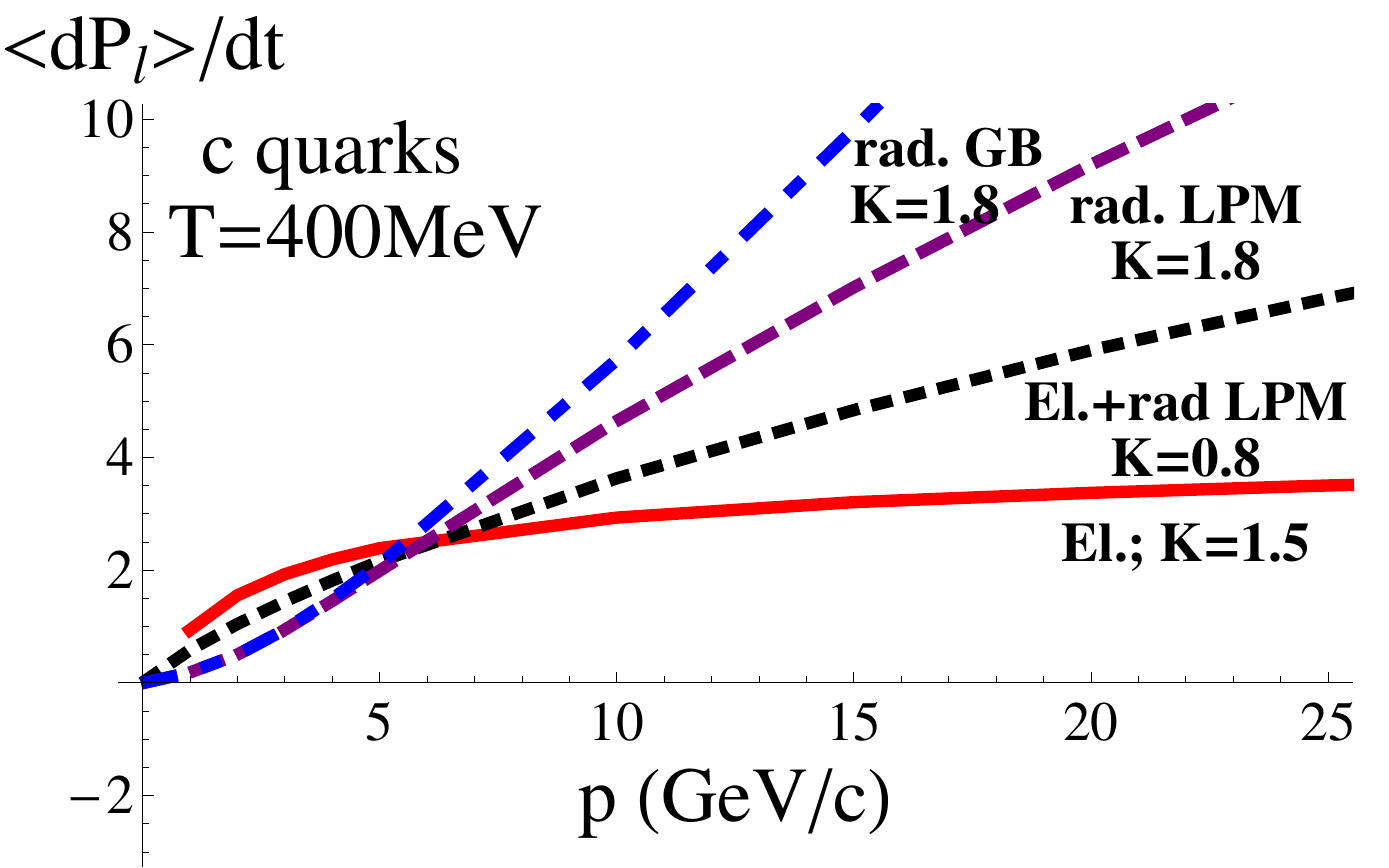}
\includegraphics*[width=7.cm]{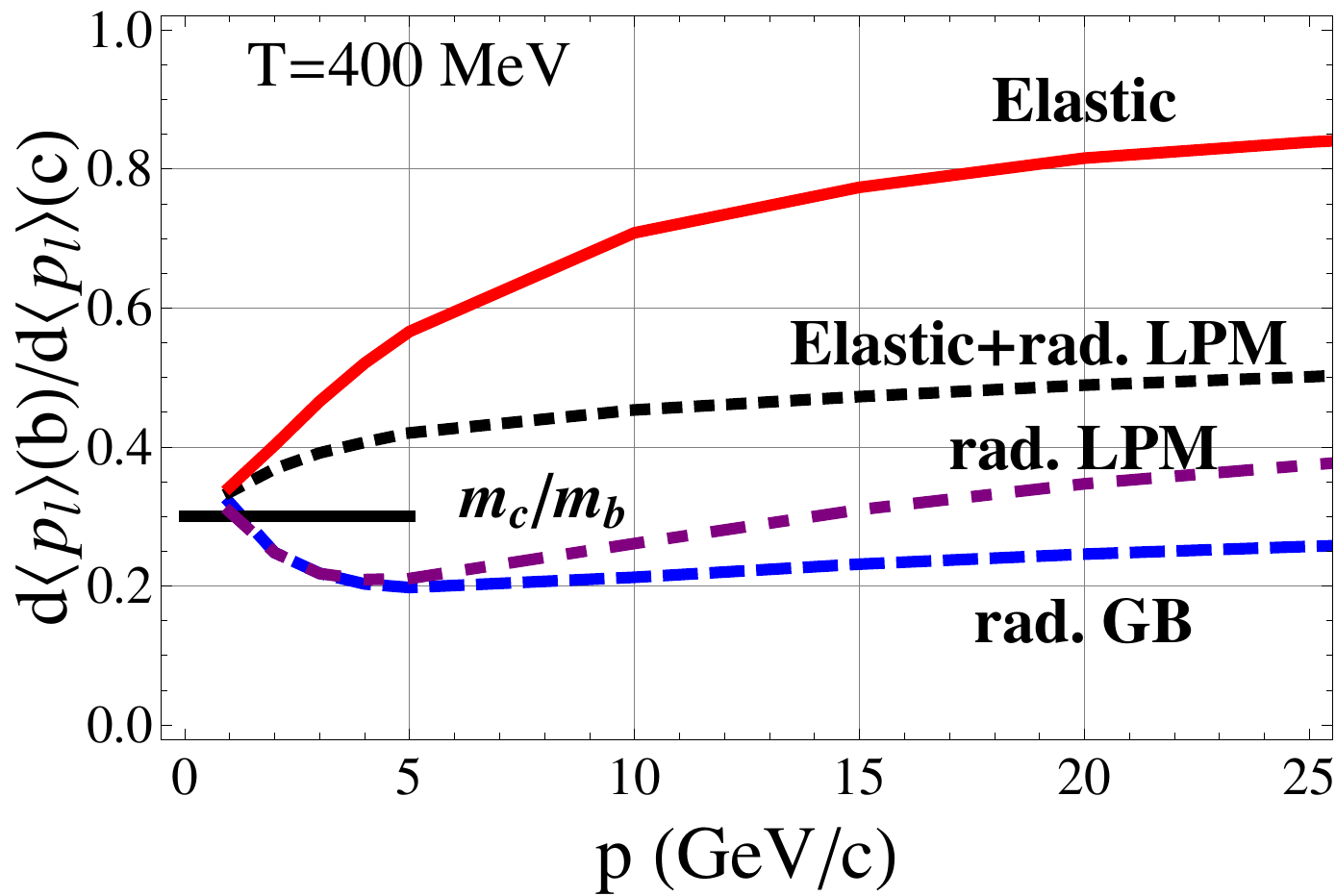}
\caption{Left: average momentum loss for different models. Right: corresponding ratio of average momentum loss of 
$b$-quarks wrt $c$-quarks.}
\label{fig:dependence_momentum_loss}
\end{center}
\end{figure}

We now discuss the mass dependence of several Eloss models implemented in MC$@_s$HQ, 
including the ones discussed in section \ref{sect_rad_eloss_interm} (rad. GB and LPM). 
On fig.~\ref{fig:dependence_momentum_loss} (left), we show the $\langle dP_l\rangle/dt$ for a $c$-quark in a 
$T=400~{\rm MeV}$ QGP. For each model, one has applied an additional factor $K$ to the interaction cross sections in order to  
reproduce the $R_{AA}$ of $D$-mesons for $p_T=10~{\rm GeV/c}$ in the most central Pb-Pb collisions at LHC
$\sqrt{s}=2.76~{\rm TeV}$ \cite{NahrgangSQM}. 
On fig.~\ref{fig:dependence_momentum_loss} (right), we show the ratio of the momentum loss of $b$-quarks as compared 
to the $c$-quarks, for the same energy-loss models. For small $p$, the ratio is close to $m_c/m_b$,
confirming the generic nature of the mass hierarchy. With increasing $p$, mass hierarchy is the strongest for the
radiative GB Eloss -- in agreement with the $m_Q^{-1}$ scaling mentioned previously -- and tends to disappear from radiative 
Eloss once coherence effects are included, while the fastest disappearance of the mass hierarchy is observed for collisional Eloss.  

\section{Consequences for heavy ion collisions}
\label{conseq_URICH}
\begin{figure}[H]
\begin{center}
\includegraphics*[width=8.cm]{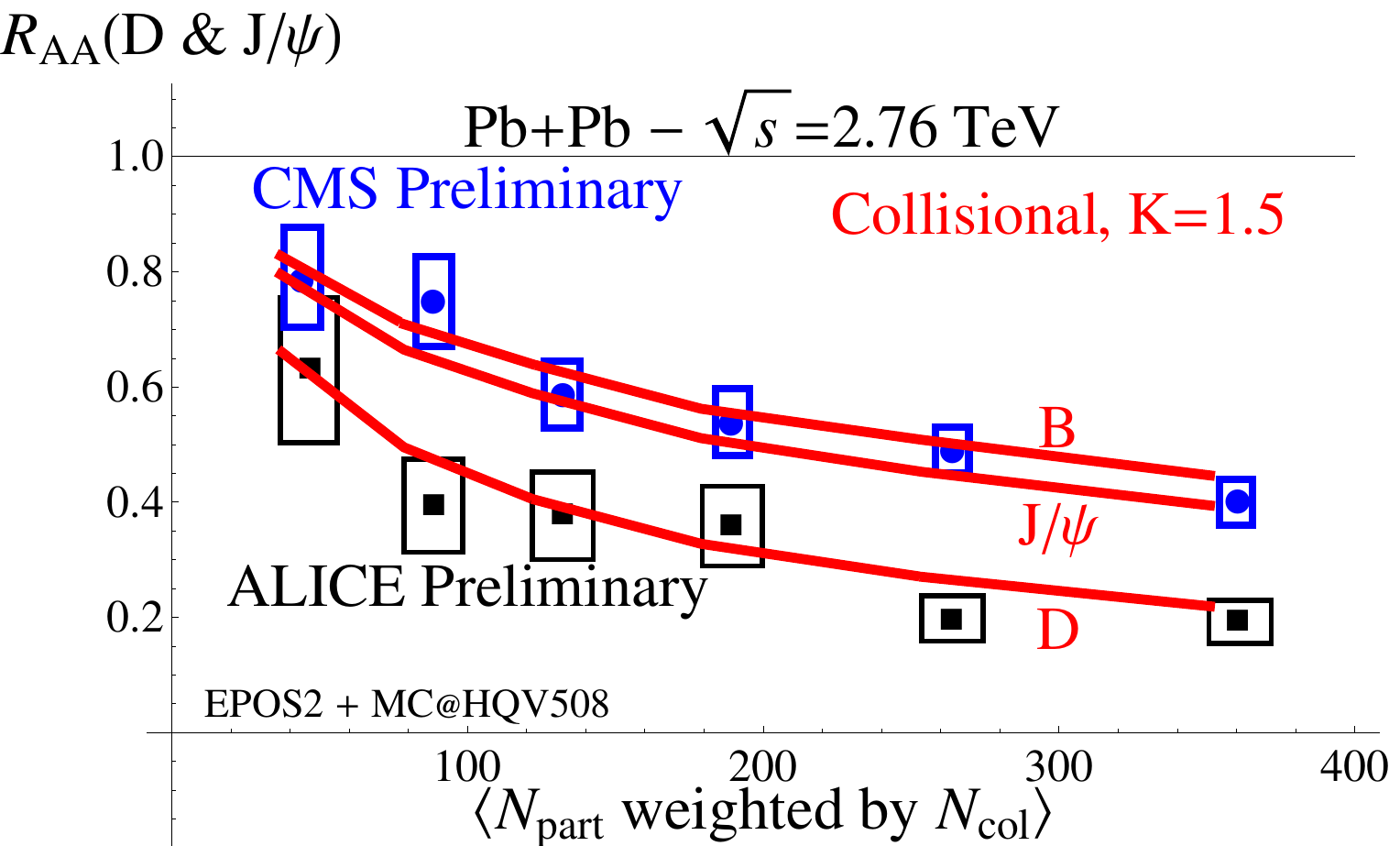}
\includegraphics*[width=8.cm]{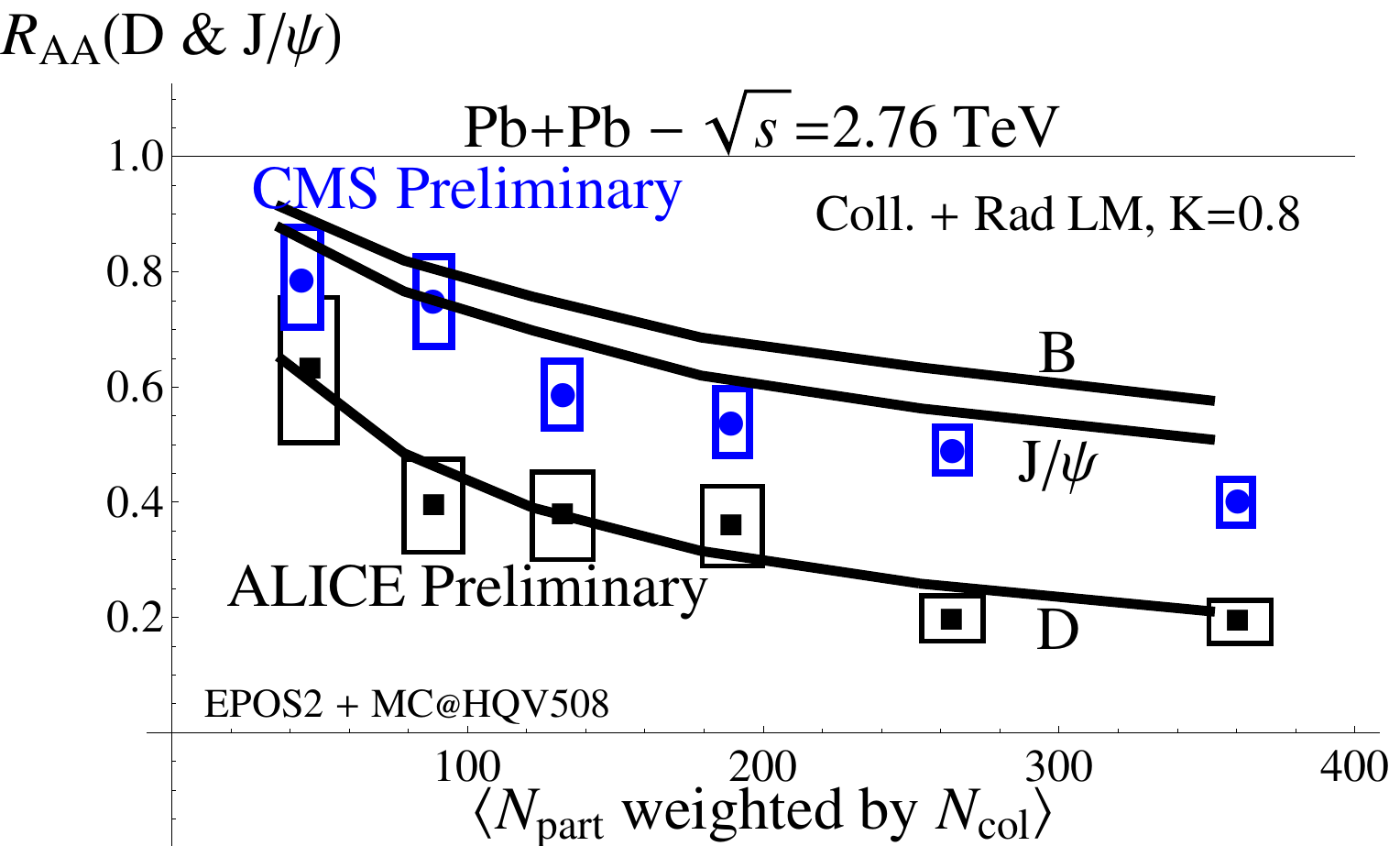}
\caption{$R_{AA}$ of $D$ mesons, $B$ mesons and non-prompt $J/\psi$ daughters, as a function of the
number of participants, for a purely collisional Eloss model (left) as well as for a
collisional + radiative cocktail (right), compared with the preliminary data of ALICE~\cite{alice} and 
CMS~\cite{CMS}.}
\label{fig:raa_of_part}
\end{center}
\end{figure}

We now investigate the consequences of the mass hierarchy on some experimental observables.
For this purpose, we implement these models in our EPOS+MC@sHQ code, described in \cite{Nahrgang1}. We first concentrate,
in fig. \ref{fig:raa_of_part}, on the comparison between the $R_{AA}$ of $D$-mesons and non-prompt $J/\psi$ stemming from $B$-mesons,
as a function of $N_{\rm part}$. Such comparison is known to be a good probe of the mass hierarchy~\cite{horrow}. 
The actual experimental data are compatible with the mass hierarchy of both the pure collisional 
scenario as well as of the collisional and radiative LPM cocktail, with a slight preference for the first type
of mass hierarchy (showing the fastest disappearance with increasing momentum). Let us also mention that
a) a pure radiative scenario is not able to cope with the combined $D$-mesons and non-prompt $J/\psi$ data while b) $b$-quark
evolution using the mass of $c$-quark leads to $B$-mesons $R_{AA}$ close to those of the $D$-mesons. This demonstrates 
that the mass ordering observed in the $R_{AA}$ is a genuine consequence of the mass hierarchy implemented in the models. 

\begin{figure}
\begin{center}
\includegraphics*[width=8.cm]{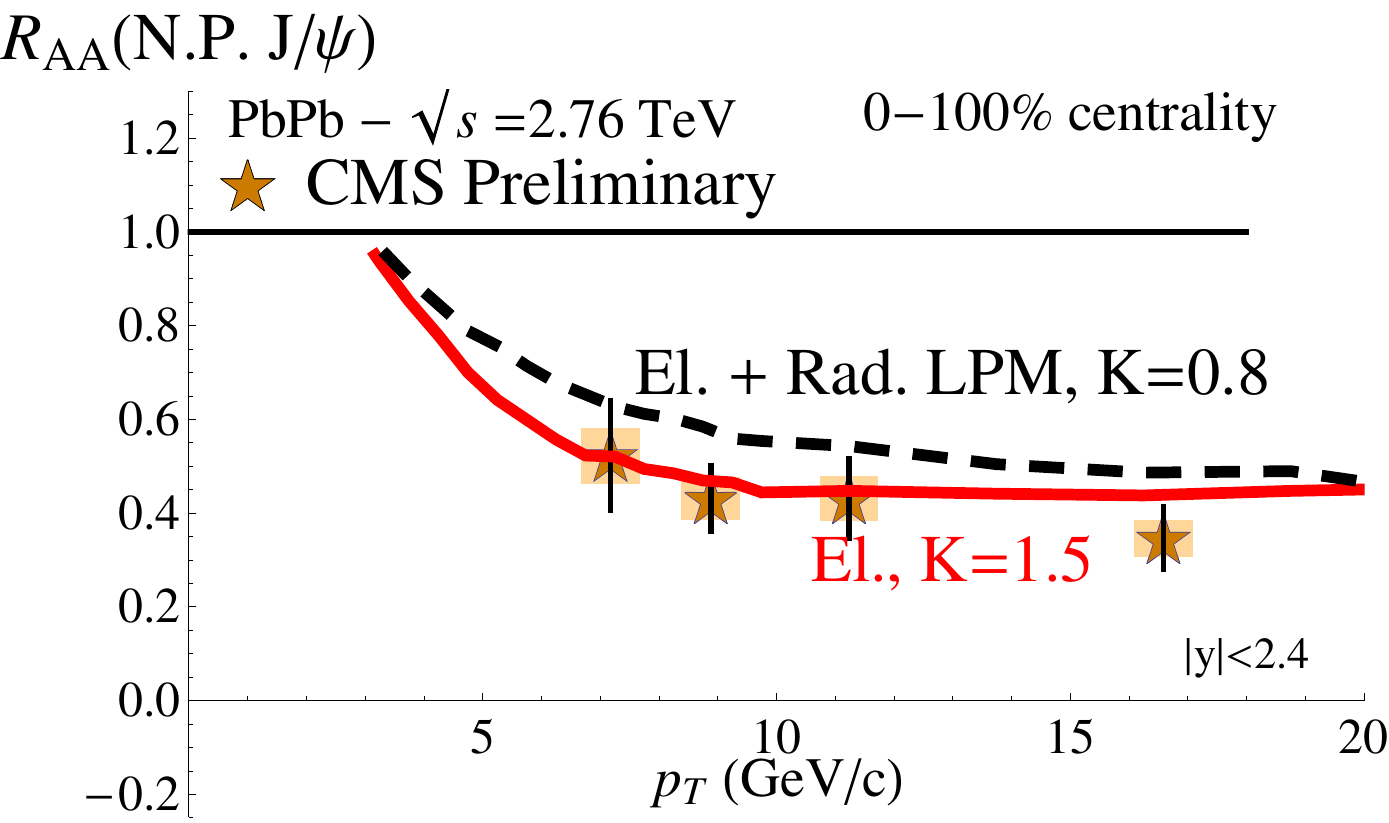}
\includegraphics*[width=8.cm]{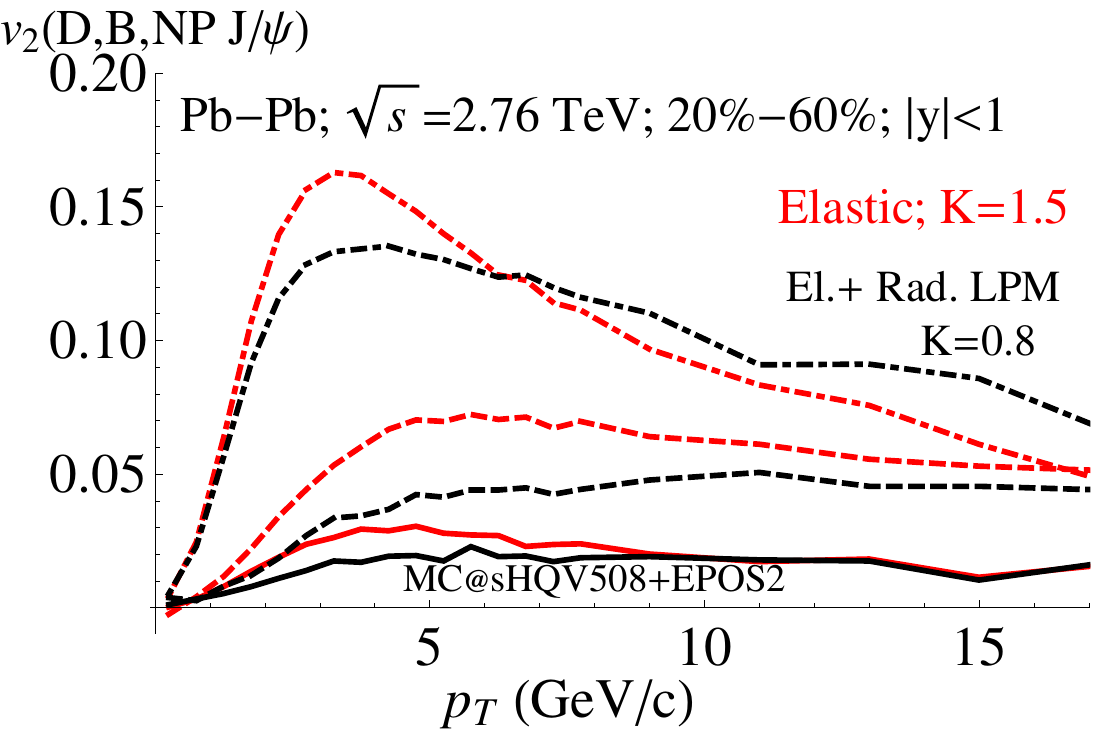}
\caption{Left: $R_{AA}$ of non-prompt $J/\psi$ (decay product of $B$-mesons) for 
pure collisional Eloss model as well as for a collisional + radiative cocktail, as compared
with CMS preliminary data. Right: prediction for the elliptic flow of $D$ (dot-dashed), $B$ (dashed) and non-prompt
$J/\psi$ mesons, using the same models.}
\label{fig:raa_of_pt}
\end{center}
\end{figure}

On fig. \ref{fig:raa_of_pt} (left), we display our prediction for the $R_{AA}$ of non-prompt $J/\psi$ vs $p_T$ in unbiased Pb-Pb
collisions at $\sqrt{s}=2.76~{\rm TeV}$. The comparison with preliminary CMS data is acceptable for the 2 types of Eloss models. 
The elliptic flow ($v_2$) of $B$ mesons, shown in the right panel of fig.~\ref{fig:raa_of_pt}, appears to be sensitive to the 
mass hierarchy as well, but is reduced when measuring the $v_2$ of the non-prompt $J/\psi$ daughters.  

\section{Conclusions}
We have contributed to the study of radiative energy loss by considering the case of heavy-quark at intermediate 
energies for which taking care of the phase space boundary appears to be a crucial issue.
We have shown that our models of the HQ interactions with the QGP medium were compatible with the $R_{AA}$ of $D$-mesons
and $B$-mesons for all centrality classes in Pb-Pb collisions at LHC, while discrimination requires further improvement in the data.
From this understanding, we were able to make a prediction for the $v_2(B)$ which will be released soon by the CMS collaboration. 

\section{Acknowledgement}
We are grateful for support from I3-Hadronphysics II, project TOGETHER (Pays de la Loire) and the U.S. department of Energy
under grant DE-FG02-05ER41367.



%% References


%\bibliographystyle{elsarticle-num}



\end{document}